\documentstyle[twocolumn,epsf,aps]{revtex}

\begin{document}
\draft
\preprint{HEP/123-qed}

\title{Anisotropic GaAs island phase grown on flat GaP: \\ spontaneously formed
quantum wire array}

\author{B. Jonas Ohlsson, Mark S. Miller\cite{byline}, and Mats-Erik Pistol}

\address{Department of Solid State Physics\\ Lund University, Box 118 S-221 00 LUND, Sweden}

\date{\today}
\maketitle

\begin{abstract}
A dense phase of GaAs wires forms in the early stages of strained growth on GaP,assembling from elongated Stranski-Krastanow islands. 
The electron diffraction during growth is consistent with long, faceted 
GaAs islands that are anisotropically deformed without dislocations.
The lateral wire period and long shapes are not predicted by published
models, though we conclude that the island orientation is picked out by facet
energy inequivalencies not present in the analogous system of Ge islands on Si.
\end{abstract}



Crystal growth on a substrate with a smaller lattice spacing can 
initially give small regular islands without crystal dislocations.
Two examples are the tens-of-monolayers-high Ge islands that form 
on Si \cite{Eaglesham} and InAs islands that form on GaAs 
\cite{Leonard1994,Gerard1995}. 
The layer-substrate system lowers its energy when the islands relax 
towards their larger lattice constant at the expense of both locally 
straining the substrate and increasing the surface area. 
The islands can behave as an equilibrium morphological phase 
\cite{Leonard1994,Miller_Madrid} with narrow size distributions around 
specific sizes \cite{Colocci1997} that often do not show appreciable coarsening.
Beyond being a recently appreciated crystal growth phenomenon, interest in 
the topic stems from the technological importance of growing strained 
semiconductor layers. 
Some combinations, such as InAs on GaAs, exhibit appreciable quantum 
confinement of electrons and holes, so the work has been further motivated by 
the study and possible application of island optical and electrical properties. 

The relative contributions of growth kinetics and equilibrium thermodynamics 
to island formation have not been settled. 
Some hypotheses for the growth kinetics selecting the island size include that
a strain-related diffusion barrier prevents atoms from moving onto 
larger islands \cite{Seifert1996,Chen1996} or that a nucleation barrier 
for additional lattice-planes on island facets limits growth \cite{Mo}. 
An equilibrium-based idea for explaining the island size and size 
distribution is that nearly uniform one-monolayer-high precursor islands 
form at equilibrium and then collapse and roll up to give the 
observed islands \cite{Priester_and_Lanoo}. 
In another approach to explain the elongated islands found in some instances, 
for example Ge islands on Si  \cite{Mo}, one model fixes the island height 
and minimizes the sum of strain and surface energy as a function of size 
and aspect ratio to give symmetric islands at an equilibrium size and
elongated islands when the islands grow beyond the preferred size \cite{TandT}. 

We report on GaAs islands grown on GaP, which can be viewed as the 
III-V semiconductor analog of Ge on Si, since the respective 
lattice constants are very close to one another.
The distinguishing behavior of this new system is that very long 
islands assemble into a dense, stable, corrugated phase with 
specific orientation and specific lateral period. 
Atomic force micrographs for samples with different GaAs 
deposition thicknesses, but otherwise made with the same growth temperatures, 
pressures and rates, show that the corrugations have appeared at a 
thickness near three monolayers and still dominate the morphology 
at approximately seven monolayers of deposition. 
The 13 nm corrugation period is in the range desired for semiconductor 
quantum wires with strong carrier confinement to one dimension. 
Electron diffraction from the corrugated phase 
shows GaAs crystals compressed to the GaP lattice constant 
in the long direction, having the GaAs lattice constant perpendicular 
to the wire in the substrate plane, and being extended beyond the 
GaAs lattice constant perpendicular to the substrate. 
The present wire phase is not found to be predicted by an 
elasticity model for strained corrugated surfaces \cite{Shchukin}, 
and we argue that the out-of-equilibrium requirement for the
long islands of Ref. \cite{TandT} are not met under the present conditions.
After considering the several reported orientations of other elongated
islands, we conclude that the intrinsic
surface stress anisotropy does not select island
orientation but that inequivalent facet energies do.
We modify the model of Ref. \cite{TandT} to account for the wire phase 
orientation by including inequivalent island facet surface energies. 

The experiments consisted of seven samples with GaAs depositions of 
0, 0.84, 1.7, 3.3, 5.0, 6.7, and 10 monolayers made in a chemical beam 
epitaxy machine of our own design.
In this growth technique, molecular beams of chemical precursors 
impinge on a heated substrate for reaction and epitaxial growth. 
For both GaAs and GaP, the growth rate is determined by the arrival 
of Ga atoms while a slight overpressure of the respective As or P 
molecular beam is maintained. 
The precursors were triethylgallium (TEGa) and thermally cracked 
{\it tertiary}-butylarsine (TBAs) and {\it tertiary}-butylphosphine (TBP). 
For all samples a 180 nm GaP buffer layer was first grown on flat 
($ \pm  0.2^{\circ}$) S-doped (001) GaP substrates at ${640}^{\circ}$C. 
The GaAs growth conditions were chosen to emphasize equilibrium behavior; 
the temperature was ${580}^{\circ}$C and the rate was 0.056 monolayers/s. 
Reflection high-energy electron diffraction (RHEED) intensity patterns 
were recorded during each growth \cite{Jag2}. 
Atomic force microscopy (AFM) measurements in tapping mode were done 
outside the vacuum system. 

\begin{figure} 
\caption{AFM micrographs for different GaAs deposition amounts. 
{\bf a.} Epitaxial GaP surface (0 monolayers of GaAs). {\bf b.} 0.84 GaAs 
monolayers showing small elongated islands. {\bf c.} Corrugated surface 
after 5.0 GaAs monolayers.  {\bf d.} Irregularly shaped islands after 10 
GaAs monolayers.}
\label{afm} 
\end{figure}

The AFM images shown in Fig. 1 follow the surface morphology with 
increasing deposition, displaying data from the 0, 0.84, 5.0, 
and 10 monolayer samples. 
Monolayer-high steps are apparent on the GaP surface of Fig.1(a), 
which was not exposed to the As molecular beam.  
At 0.84 monolayers of GaAs, Fig. 1(b), islands have formed with heights 
of up to 6 monolayers, or 2 nm. 
The islands do not exhibit a sharply defined size or shape, though 
there is a clear tendency for elongation in the [1\={1}0] direction. 
The widths in the [110] direction are 20 to 27 nm, and many islands 
have in-plane aspect ratios close to 4:1. 
The density of islands is 1x$10^{10}$ cm$^{-2}$.  
The density and length of the islands increase with deposition until 
condensing into the corrugated phase shown in Fig. 1(c) for the 5.0 
monolayer deposition sample. 
The corrugation spacing varies from 10 to 15 nm. 
Many of the GaAs wire structures can be seen to extend for more 
than 0.25 $ \mu$m. 

\begin{figure}[tbp]  \epsfxsize=8.5cm\epsfbox{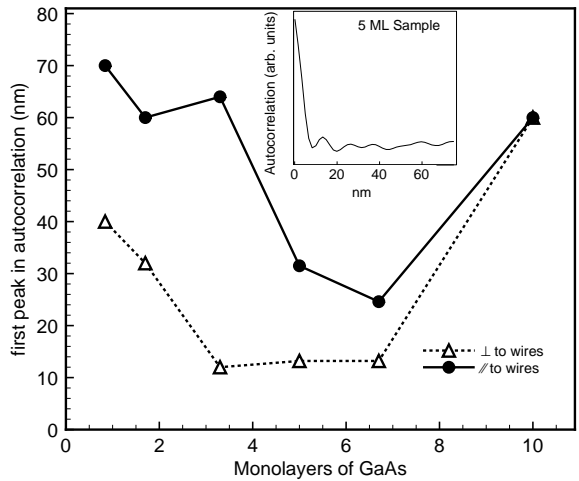}
\caption{ The first autocorrelation peak of the surface feature heights 
as a function of deposited amount of GaAs. 
The peak is a measure of the separation of the nearest neighbor corrugations. 
The inset shows the autocorrelation from the 5.0 monolayer sample taken 
perpendicular to the wires.}
\label{autocorrelation} \end{figure}

The measured heights of the corrugations are close to 2 nm. 
If the actual island heights are greater, the close lateral spacing 
could be preventing the AFM tip from reaching the substrate between them. 
Though not shown in Fig. 1, the corrugations have appeared and already 
cover approximately 90 \% of the surface for the 3.3 monolayer deposition. 
The corrugations similarly dominate the morphology of the 6.7 monolayer 
sample,  also not shown here, though irregular compact islands have 
begun to form. 
Larger, irregular compact islands with heights of up to 13 nm cover 
the 10 monolayer sample of Fig. 1(d).

Figure 2 summarizes autocorrelations of AFM feature heights taken for all 
of the samples. 
Autocorrelation was chosen to facilitate the comparison of island 
separations in non-periodic island morphologies with those in the 
approximately periodic corrugations. 
Height autocorrelations were calculated for each row (column) of data 
making up the image and then summed to give 1D average autocorrelations 
parallel (perpendicular) to the wires. 
As an example, the result for the corrugated surface of Fig. 1(c) 
perpendicular to the wires is given in the inset of Fig. 2. 
The approximate periodicity in the inset reflects the approximate 
periodicity of the corrugations. 
The first peak in the autocorrelation is a measure of the average 
nearest neighbor separation of the islands parallel and perpendicular 
to the wires and is plotted in Fig.2 for all samples. 
The 40 nm and 70 nm peaks of the 0.84 monolayer sample reflect the 
$10^{10}$ cm$^{-2}$ density of isolated islands, where the asymmetric 
values are due to the elongated island shapes. 
The [110] island-island separation decreases with deposition until 
the appearance of the corrugations near 3 monolayers, where it reaches 
13 nm and remains constant until around 7 monolayers. 
Throughout the corrugated phase the peak in the [1\={1}0] direction 
decreases as the number of compact irregularities along the wires increases. 
At 10 monolayers the first peaks are at 60 nm for both the  [1\={1}0] and 
the [110] directions, reflecting the statistical isotropy of both 
shape and positional correlations of the large irregular islands.

\begin{figure}[tbp]  \epsfxsize=8.5cm\epsfbox{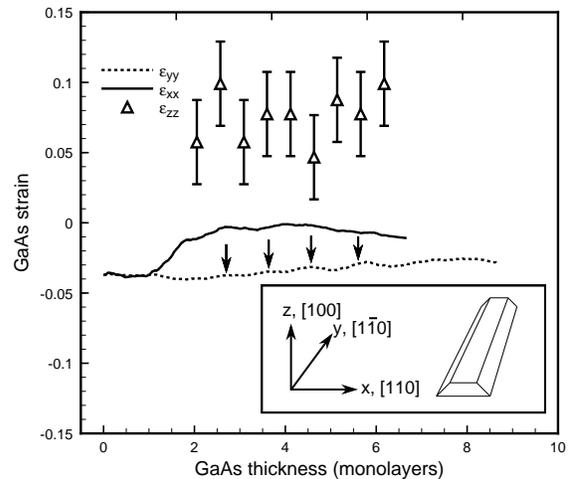}
\caption{Diagonal strain tensor components of the GaAs material measured 
with RHEED and plotted as a function of GaAs layer thickness for three 
different crystal directions. 
The inset depicts the strain component orientations with respect to 
the islands.}
\label{Strain} \end{figure}

The RHEED patterns recorded during growth give a measure in reciprocal
space of the evolving surface morphology \cite{Jag2}.
When diffraction was recorded with the electron beam perpendicular to the
wires, the diffraction pattern changed continuously from the
reflection-diffraction rods of a 2-fold reconstructed surface to
the transmission-diffraction spots of the GaAs islands.
With the beam parallel to the wires, a similar continuous transition was
recorded from an initial 4-fold pattern, though chevron-shaped features 
at the transmission spots were also seen.
The chevron angle of approximately $ {35}^{\circ} $ would correspond to
island facets along the wires of {114} orientation.

The position of the diffraction features depends on the island strain.
Estimates of the diagonal strain tensor components of the islands are 
plotted versus deposition thickness in Fig. 3, where GaP rod separations were
used as references.
For strains parallel to the surface, $ {epsilon}_{xx} $ 
perpendicular to the wires and ${\varepsilon}_{yy} $ along the wires, the 
(\={1}0) to (10) rod separation was measured at positions corresponding 
to 3D diffraction. 
Transmission spot separations perpendicular to the surface gave the
strain perpendicular to the surface,  $ {\varepsilon}_{zz} $.
Along the wires, in the [1\={1}0] direction, the islands stay strained to 
close to the GaP substrate throughout the recorded 10 monolayers of deposition.
In the [110] direction, perpendicular to the wire axis, the lattice constant 
begins to relax to that of GaAs after approximately one monolayer, once 
the islands have formed and are contributing a significant amount of 3D 
diffraction intensity. 
By 3 or 4 monolayers, the sides of the islands have bulged out to 
close to the GaAs lattice constant. 
The [001] tensile island strain perpendicular to the surface in the 
corrugated phase is $ \sim$0.06. 
Taken together, the strain measurements in the three perpendicular 
directions indicate that the long islands in the corrugated phase are 
elastically deformed, implying no crystal dislocations.
The present estimates treat the diffraction feature positions as giving
a measure of uniformly strained islands.
A more accurate treatment would model the nonuniform strain in the islands,
the non-uniform penetration of the electron beam into the islands, and
dynamical scattering processes during diffraction.
The Arrows in Fig. 3 points out oscillations as function of monolayer completion, wich have been reported previously for the InAs/GaAs system by \cite{Massies}. 
More detailed RHEED analysis can be found in Ref. \cite{Jag2}.

The period found for the dense GaAs wire phase reported here can be 
compared with a continuum elasticity model that relates lattice mismatch 
with surface corrugation period \cite{Shchukin}. 
In that model, a local minimum was found in the Helmholtz free energy, given 
by $ {E}_{T}={E}_{facets}+{E}_{edges}+{E}_{elastic} $. 
Here $ {E}_{facets} $ is the free energy of the facets, $ {E}_{edges} $ is 
an energy associated with the facet edges and $ {E}_{elastic} $ is an elastic 
strain energy due to edges and lattice mismatch. 
A local energy minimum corresponding to the corrugated surface was found only
to exist for a lattice mismatch that is less than a critical value 
determined by the corrugation size and material parameters. 
The critical mismatch is   
\begin{eqnarray}
{\left( \Delta a \over a \right)}_{c} =
c{{\tau}_{xx} \over {L}_{0}},
\end{eqnarray}
where $ {\tau}_{xx} $ is the facet intrinsic surface stress perpendicular 
to the edges, $  {L}_{0}$ is the period of the faceting, and $c$ is a 
constant dependent on the elastic moduli and geometry. 
The present GaAs corrugated surface has $ {L}_{0}=13$ nm.  
Here we use the same approximate value of  $ {\tau}_{xx}=100 $ 
meV/ $ {{\rm \AA}}^{2}$ as in Ref. \cite{Shchukin} and find
a critical lattice mismatch of 
${\left( \Delta a \over a \right)}_{c}=2.3 \times {10}^{-3} $. 
This upper bound is an order of magnitude smaller than the present 
GaAs/GaP lattice mismatch of 
$\left( \Delta a \over a \right)=3.7\times{10}^{-2} $. 
Although the corrugated surfaces treated by the model 
have obvious similarities to the present GaAs surface, the large mismatch 
and short period mean that this model can not be directly applied.

The shape of the long narrow islands of GaAs on GaP can be compared 
with a model for long rectangular islands.
The island energy was taken as
${E}_{T}={E}_{facets}+{E}_{elastic}$ \cite{TandT}, 
and changes in size and in-plane aspect ratio were considered.
For a fixed island height, an equilibrium size was found, and the shape
of this optimal island was a symmetric square.
However, if islands larger than the equilibrium size were considered,
then the optimal shape was found to be asymmetric with approximately
constant width but increasing length with size.
It was suggested that isolated islands become over ripe 
during growth when they are too far apart to exchange atoms through diffusion. 
When the height is not fixed, the square island always has the lower 
energy \cite{Khor}. 
Consistent with the model requirements, the height of the GaAs islands 
remains approximately constant when they form at low coverage and 
throughout the corrugated phase. 
In accord with the model prediction, the width of the islands remains 
approximately constant while their length increases, apparently only 
limited by irregularities. 
Still, the 13 nm corrugations are small enough that islands should 
easily exchange atoms through diffusion. 
For comparison, appreciable diffusive mass transport occurs over distances 
greater than 100 nm during GaAs growth on lithographically patterned substrates.
At our slow deposition rate and warm temperature, the islands should easily 
exchange atoms and the model would predict symmetric islands. 
Thus, the corrugated surface represents an optimal 
equilibrium shape in itself without any over ripening. 
The result of Ref. \cite{TandT} does not account for the shape of these islands
because the predicted shape would be symmetric for equilibrium conditions. 

The orientation of the wires might be determined by several contributions.
One possibility is the strongly anisotropic surface stress which is aligned 
with the dimers of the reconstructed (001) surface.
On Si,  where the dimer orientation rotates by ${90}^{\circ}$ with each 
atomic surface step, the elastic interaction is sufficient to spontaneously 
rearrange a flat surface into alternating domains \cite{Alerhand}.
Here we make orientation comparisons with other elongated island systems.
The analogous Ge islands on Si are 2 to 4 nm high, 25 nm wide, and can 
be over 100 nm long \cite{Mo}. 
Despite the similar dimensions, the Ge islands grow with
two orientations, along both [010] and [100], which are ${45}^{\circ}$ to 
the present [1\={1}0] oriented GaAs islands.  
In another III-V system, InSb islands on InP (10.4\% mismatch) show 
a shape transition with increasing size to elongation along [110], 
where the islands are 10 nm high, 30 nm wide, and 90 nm long \cite{Utzmeier}.  
As a closer mismatch analog (3.2\%) having the same column-V elements, 
InAs islands formed on InP substrates show a shape transition with larger 
symmetric islands transforming to smaller elongated islands.
The InAs islands on InP are 3 to 4 nm high, 30 nm wide, over 100 nm long, 
and aligned in the [1\={1}0] direction \cite{Ponchet}. 
For both the III-V and IV semiconductors, the dimers 
are parallel to $\left\langle110\right\rangle $ directions. 
The III-V materials typically have V-terminated surfaces with dimer 
axes parallel to the [1\={1}0] direction. 
Because islands with many orientations with respect to the intrinsic
surface stress field are reported, we conclude that the stress anisotropy 
does not play the determining role in island orientation.

The difference in facet surface energies may be picking the island orientation.
The Ge island facet energies islands will reflect the cubic crystal symmetry, 
and consistent with this two perpendicular orientations are found.
The facet energies of the polar III-V compounds will reflect the lack of 
inversion symmetry, with the largest difference between III- and V-terminated 
$ \left\{111 \right\}$ surfaces. 
While the model of Ref. \cite{TandT} does not account for long equilibrium
shapes, the model can be modified to account for inequivalent orientations by
introducing two facet surface  energies, $ {E}_{A} $ and $ {E}_{B} $.
The the total energy divided by the island volume is then

\begin{eqnarray}
{E \over V}={2 \over  \sin{\theta}}\left({ {E}_{A} \over t}  +{ {E}_{B} \over s} \right)+{E}_{s}
\cot{\theta} ({s}^{-1} +{t}^{-1})
\nonumber \\
-2ch \left( {s}^{-1}\ln{{s \over h\phi}} +{t}^{-1}\ln{{t \over h\phi}} \right),
\end{eqnarray}

with $ {E}_{s} $ the top facet surface energy; {\it c} a constant containing
elastic moduli; {\it s}, {\it t}, and {\it h} the two island side lengths 
and the height; and $ \phi $ a constant containing the facet angle $ \theta $.
The inequivalent facet surface energies that appear in the first term
serve to provide the orientational bias.
As expected, the longer island side will have the lower surface energy.

To conclude, a dense GaAs wire phase was found to form on GaP.  
Diffraction patterns show elastically strained islands with diffraction 
features consistent with reflection from island facets. 
The long GaAs islands are similar to Ge islands found on Si. 
Though a crucial difference between the two systems may be that the 
GaAs wires grow only in one orientation, allowing the condensation into a dense
ordered phase. 
The orientation of elongated islands was deduced to be due to differences in 
facet energies rather than to surface stress anisotropies. 
After comparison with available theories, open questions include:  
What picks the size and what gives the long shapes?

Tobias Junno and Jonas Tegenfeldt are thanked for helpful discussions 
concerning AFM imaging. 
Martin Magnusson is thanked for critical reading of the manuscript. 
This work was
performed within the Nanometer Structure Consortium in Lund.

\end{document}